\documentclass[12pt]{article}
\usepackage{graphicx}
\usepackage{amsmath}
\usepackage{accents}
\usepackage{relsize}

\newcommand\pubnumber{}
\newcommand\pubdate{}

\def\Title#1{\begin{center} {\Large #1 } \end{center}}
\def\Author#1{\begin{center}{ \sc #1} \end{center}}
\def\Address#1{\begin{center}{ \it #1} \end{center}}

\newcommand\pubblock{\rightline{\begin{tabular}{l} \pubnumber\\
         \pubdate  \end{tabular}}}
\newenvironment{Abstract}{\begin{quotation}  }{\end{quotation}}
\newenvironment{Presented}{\begin{quotation} \begin{center} 
             PRESENTED AT\end{center}\bigskip 
      \begin{center}\begin{large}}{\end{large}\end{center} \end{quotation}}

\def\qoverp {\ensuremath{\frac{q}{p}}\xspace}
\def\poverq {\ensuremath{\frac{p}{q}}\xspace}
\def\iotwo {\ensuremath{\frac{i}{2}}\xspace}
\input babarsym
\usepackage{relsize}
\def\babar{\mbox{\slshape B\kern-0.1em{\smaller A}\kern-0.1em
    B\kern-0.1em{\smaller A\kern-0.2em R}}}
\def\superb{Super\!$B$}

\begin{document}
\begin{titlepage}
\pubblock

\vfill
\Title{Quantum Correlated $D$ Decays at \superb}
\vfill
\Author{ Chih-hsiang Cheng}
\Address{California Institute of Technology, Pasadena, California 91125, USA}
\vfill
\begin{Abstract}
  We present the prospects for studying quantum correlated charm decays at the
  $\psi(3770)$ using 0.5--1.0\invab of data at \superb.  The impact of studying 
  such double tagged decays upon measurements in other
  charm environments will be discussed.
\end{Abstract}

\vfill
\begin{Presented}
The $5^{\rm th}$ International Workshop on Charm Physics\\
Honolulu, Hawaii, USA, May 14--17, 2012
\end{Presented}
\vfill
\end{titlepage}
\def\thefootnote{\fnsymbol{footnote}}
\setcounter{footnote}{0}

\section{Introduction}

\superb\ is a next-generation high-luminosity asymmetric-energy $e^+e^-$ 
collider that 
aims to collect 50 to 100 times more data than the $B$-Factories of the
last decade, \babar\ and Belle. The center-of-mass energy for the majority
of the program will be at or near the $\Upsilon(4S)$ resonance with the
designed peak luminosity of $10^{36}\ {\rm s}^{-1}{\rm cm}^{-2}$. The goal 
is to collect 75\invab over five years. Additional runs are also planned
at $D\Dbar$ threshold $\psi(3770)$ to collect 0.5--1.0\invab over a few
months. \superb's physics programs include, but not limited to, 
heavy-flavor $B_{u,d,s}$, $D$, and $\tau$ physics. \superb\ will be able to 
search for new physics at energy scale up to 10--100~TeV through 
rare/forbidden decay searches, \CP violation, and precision CKM matrix 
measurements~\cite{O'Leary:2010af}.

With 75\invab of data near $\Upsilon(4S)$, ${\cal O}(10^{11})$ charm mesons
will be created. They come from continuum production $\ee\to c\bar c$, as
well as $B$ decays. Many charm analyses identify a $D$ meson through 
$D^{*+}\to \Dz\pi^+$ process. Consequently the reconstruction efficiencies 
are relatively low.

With 0.5\invab of data at $\psi(3770)$, one can expect approximately
$1.8\times 10^9$ $\Dz\Dzb$ and $1.5\times 10^9$ $\Dp\Dm$ events. This 
amount is more than an order of magnitude larger than the current
charm factory BESIII~\cite{Asner:2008nq} will collect. 
As a $J^{PC}= 1^{--}$ state, 
$\psi(3770)\to D\Dbar$ is in a quantum entangled, anti-symmetric state.
If one $D$ decays to state $\alpha$ at time $t_1$ and the other to $\beta$ 
at $t_2$, the decay amplitude ${\cal M}$ is
\begin{equation}
{\cal M} = \frac{1}{\sqrt{2}}\big[ \langle\alpha|H|\Dz(t_1)\rangle \langle\beta|H|\Dzb(t_2)\rangle - \langle\beta|H|\Dz(t_2)\rangle \langle\alpha|H|\Dzb(t_1)\rangle \big].
\end{equation}
A neutral meson mixing system can be described by a $2\times 2$ effective 
Hamiltonian with non-vanishing off-diagonal terms
\begin{equation}
i\frac{\partial}{\partial t} \left (\begin{array}{c} \Dz(t) \\ \Dzb(t) \end{array} \right) = \left( {\mathbf M} - \iotwo {\mathbf \Gamma} \right)
\left (\begin{array}{c} \Dz(t) \\ \Dzb(t) \end{array} \right).
\end{equation}
The eigenstates $|D_{1,2}\rangle = p |\Dz\rangle \pm q |\Dzb\rangle$ satisfy
\begin{equation}
\qoverp = \sqrt{\frac{M_{12}^* - \iotwo \Gamma^*_{12}}{M_{12} - \iotwo \Gamma_{12}}}, \quad |p|^2+|q|^2 = 1.
\end{equation}
The eigenvalues are
\begin{equation}
\lambda_{1,2} \equiv m_{1,2}-\iotwo \Gamma_{1,2}= \left(M-\iotwo\Gamma\right)\pm\qoverp\left(M_{12}-\iotwo\Gamma_{12}\right).
\end{equation}
Here we have assumed \CPT conservation. The time evolution of $\psi(3770)\to D \Db \to \alpha(t_1)\beta(t_2)$ system can then be expressed as
\begin{align}
d\Gamma/dt \propto (|a_+|^2 + |a_-|^2 ) \cosh( y\Gamma\Delta t ) + (|a_+|^2-|a_-|^2) \cos(x\Gamma\Delta t) \nonumber \\
- 2 {\cal R}e(a_+^* a_-) \sinh(y\Gamma \Delta t) + 2 {\cal I}m(a_+^* a_-) \sinh(x\Gamma \Delta t),
\label{eq:dGammadt}
\end{align}
where $\Delta t=t_2-t_1$, $a_+\equiv \bar A_\alpha A_\beta - A_\alpha \bar A_\beta$,
$a_-\equiv - \frac{q}{p}\bar A_\alpha \bar A_\beta + \frac{p}{q}A_\alpha A_\beta$,
$M=(M_{11}+M_{22})/2$
$\Gamma= (\Gamma_{11}+\Gamma_{22})/2$, $x= (m_1-m_2)/\Gamma$, and
$y=(\Gamma_1-\Gamma_2)/(2\Gamma)$; $A_x(\bar A_x)$ is the decay amplitude of $D$ 
(\Db) to $X$.

\section{Charm mixing measurements}

The mixing in neutral $D$ system is expected to be very small. 
The short-distance
$|\Delta F|=2$ comes from box diagrams. The diagrams with $b$ quark in the loop
is CKM-suppressed (with $V_{ub}$ in the vertex),
and the ones with $s$ and $d$ quarks
are GIM-suppressed. The long-distance contributions come from diagrams that
connect $D^0$ and \Dzb through on-shell states (e.g., $K\Kb$). These 
long-distance contributions are expected to be ${\cal O}(10^{-3})$ but the
theoretical calculation is difficult~\cite{Burdman:2003rs}. 
\CP violation induced by mixing is 
therefore expected very small too. Observations of large mixing and/or \CP\ 
violation are considered clear signs of new physics beyond the standard model.

Charm mixing has been firmly established at $B$-factories~\cite{Amhis:2012bh} using
continuum events in the data taken near the $\Upsilon(4S)$ resonance.
Both
$x$ and $y$ terms are approximately 0.5\%. These analyses use the charge of
the soft pion from $D^{*+}\to\Dz\pi^+$ (or its charge conjugate process)
to identify the initial flavor of the $D$ meson, and reconstruct a final state that
is accessible by both \Dz and \Dzb. The decay time distribution for the
$D$ meson tagged by $\pi^+$ ($N(t)$) and $\pi^-$ ($\bar N(t)$) are
\begin{eqnarray}
N(t)&\propto& \left[ 1+ \frac{x^2+y^2}{4}|\lambda_f|^2 (\Gamma t)^2+
|\lambda_f|( y\cos (\overline{\delta_f+\phi_f}) - x \sin(\overline{\delta_f+\phi_f})(\Gamma t)\right] \\
\bar N(t)&\propto& \left[ 1+ \frac{x^2+y^2}{4}|\lambda_f|^{-2} (\Gamma t)^2+
|\lambda_f|^{-1}( y\cos (\overline{\delta_f-\phi_f}) - x \sin(\overline{\delta_f-\phi_f})(\Gamma t)\right] ,
\end{eqnarray}
where $x\Gamma t,\ y\Gamma t \ll 1$, $\lambda_f= (q\bar A_f)/(p A_f)$, and $\phi_f=\psi_f+\phi_m$, where $\delta_f$ ($\psi_f$) is the relative strong (weak) phase in decay, and $\phi_m$ is the mixing phase arg($q/p$).

The cleanest mode used in this method is $D\to \Kp\pim$, which is 
Cabibbo-favored in \Dzb decays but doubly-Cabibbo suppressed in \Dz decays.
One does not measure $x$ and $y$ directly. Rather, the
observables are rotated by the strong phase difference $\delta_{K\pi}$:
\begin{eqnarray}
x^\prime &=& x \cos\delta_{K\pi} + y \sin\delta_{K\pi} \\
y^\prime &=& y \cos\delta_{K\pi} - x \sin\delta_{K\pi}.
\end{eqnarray}
Independent measurements of strong phase difference are needed. 

Strong phase differences can be measured in $\psi(3770)\to D\Db$ decays with
a ``double-tag'' technique. Due to the quantum-entangled nature of the system,
when one $D$ decays to a \CP\ final state, the other $D$ is projected to the
orthogonal state, which is a linear combination of \Dz and \Dzb, and its decay
branching fraction is sensitive to the relative strong phase of $\Dz\to f$
and $\Dzb\to f$. For example, for $f= K^-\pi^+$, the effective branching 
fraction of the double-tag event is~\cite{Asner:2008ft},
\begin{equation}
{\cal F}_{S_\pm,\Km\pip} \simeq {\cal B}_{S_\pm} {\cal B}_{\Km\pip}
( 1 \pm 2r\cos\delta_{K\pi} + R_{\rm WS} + y),
\end{equation}
where ${\cal B}_{S_\pm}$ and ${\cal B}_{\Km\pip}$ are the branching fractions 
of \Dz\ decaying to $\CP\pm$ and $\Km\pip$ final states, respectively,
$\langle \Kp\pim | \Dz \rangle / \langle \Kp\pim | \Dzb \rangle = r e^{-i\delta_{K\pi}}$, and $R_M$ is the wrong-sign total decay rate ratio, $R_M\equiv\Gamma(\Dzb\to\Km\pip)/\Gamma(\Dz\ra\Km\pip)= r^2+ry^\prime+ (x^2+y^2)/2$.
CLEO-c~\cite{Asner:2008ft} has demonstrated this technique with 281\invpb of
data and obtained $\delta_{K\pi}= (22^{+11+\ 9}_{-12-11})^\circ$ or $[-7^\circ,+61^\circ]$ interval at 95\% confidence level.

Another powerful method of measuring \Dz-\Dzb mixing is using a time-dependent
Dalitz-plot analysis with three-body decays. With this method, one can avoid
strong phase ambiguity and resolve $x$ and $y$ by exploiting strong phase 
variation and interferences of resonances on the Dalitz plot.
The most power mode of this kind is $\Dz\to\KS\pip\pim$. The time-dependent
decay amplitude of a state created as \Dz or \Dzb at $t=0$ can be expressed 
as~\cite{Asner:2005sz},
\begin{eqnarray}
{\cal M}(s_{12},s_{13},t) &=& A_D(s_{12},s_{13})\frac{e_1(t)+e_2(t)}{2}+\qoverp \bar A_D(s_{12},s_{13})\frac{e_1(t)-e_2(t)}{2}, \\
\bar {\cal M}(s_{12}, s_{13},t) &=& \bar A_D(s_{12},s_{13})\frac{e_1(t)+e_2(t)}{2}+\poverq  A_D(s_{12},s_{13})\frac{e_1(t)-e_2(t)}{2}, 
\end{eqnarray}
where $A_D$ ($\bar A_D$) is the decay amplitude of \Dz (\Dzb) as a function of 
invariant mass squared $s_{12}\equiv m^2_-= (p_{\KS} + p_{\pim})^2$,
$s_{13}\equiv m^2_+= (p_{\KS} + p_{\pip})^2$,
 and 
$e_{1,2}(t) = \exp[-i(m_{1,2}-i\Gamma_{1,2}/2)t]$.
Using this method, Belle~\cite{Abe:2007rd} and \babar~\cite{delAmoSanchez:2010xz} 
measured $x= (0.80\pm 0.29^{+0.09+0.10}_{-0.07-0.14})\%$, 
$y=(0.33\pm0.24^{+0.08+0.06}_{-0.12-0.08})\%$, and 
$x=(0.16\pm 0.23\pm 0.12\pm 0.8)\%$, $y= (0.57\pm 0.20\pm 0.13\pm 0.07)\%$,
respectively, where the first uncertainties are statistical, the second are
systematic, and the third are Dalitz plot model uncertainty. 

With 75\invab of at $\Upsilon(4S)$ at \superb, the statistical uncertainty
can be reduced by a factor of 10. Since major systematic uncertainties
are in fact statistical in nature, estimated from data control samples and
simulated events, they will also be improved with more data. However, the 
Dalitz plot model uncertainty may not improve much without other input;
it will become the dominant uncertainty at \superb~\cite{O'Leary:2010af}.

To avoid Dalitz plot model dependence, Giri  {\it et al}~\cite{Giri:2003ty} proposed
a method, originally for measuring the CKM angle $\gamma$ in 
$\Bp\to D[\KS\pip\pim]\Kp$ decays using time-dependent Dalitz plot analysis.
In this method, the Dalitz plot phase space is divided into $N$ pairs of bins;
two bins in each pair is mirror-symmetric about the line $s_{12}=s_{13}$
the Dalitz plane. One then can define
\begin{eqnarray}
c_i &\equiv& \int_i dp A_{12,13}A_{13,12} \cos(\delta_{12,13}-\delta_{13,12}), \\
s_i &\equiv& \int_i dp A_{12,13}A_{13,12} \sin(\delta_{12,13}-\delta_{13,12}), \\
T_i &\equiv& \int_i dp A^2_{12,13},
\end{eqnarray}
where $\delta_{1j,1k}\equiv\delta(s_{1j},s_{1k})$, and $A_{1j,1k}$ is the magnitude
of the $D$ decay amplitude $A_D(s_{1j},s_{1k})=A_{1j,1k} \exp(i\delta_{1j,1k})$.
The integral is over the phase space of the bin $i$. 
Here we have used the fact that $A_D(s_{12},s_{13})= \bar A_D(s_{13},s_{12})$.
The $c_i$ and $s_i$ contain unknown strong phase difference $\delta_{12,13}-\delta_{13,12}$, and thus unknown, but $T_i$ can be measured with flavor tagged $\Dz$ decays. For mirror bins, $i$ and $\bar i$, $c_i=c_{\bar i}$ and
$s_i= -s_{\bar i}$.
With charm mixing, the number of events in bin $i$ at time $t$ 
is~\cite{Bondar:2010qs}
\begin{equation}
T^\prime_i(t) \propto e^{-\Gamma t} [ T_i + \sqrt{T_iT_{\bar i}}(c_i y + s_i x)\Gamma t
+ {\cal O}((x^2+y^2)(\Gamma t)^2)].
\end{equation}
One can fit all bins simultaneously to extract mixing parameters $(x, y)$ if
$(s_i, c_i)$ are known. 

Again, using entangled $\psi(3770)\to D\Db$, one can measure $s_i$ and $c_i$.
If one $D$ decays into a \CP eigenstate, the other $D$ is in an orthogonal
state. We denote these two states as $D_{\pm}^0 \equiv (\Dz\pm\Dzb)/\sqrt(2)$.
The amplitude and partial decay width of the second $D$ can be written 
as~\cite{Giri:2003ty}
\begin{multline}
\quad 
A(D_{\pm}^0\to \KS(p_1)\pim(p_2)\pip(p_3))=\frac{1}{\sqrt{2}}(A_D(s_{12},s_{13})
\pm A_D(s_{13},s_{12})),\\
d\Gamma(D_{\pm}^0\to \KS(p_1)\pim(p_2)\pip(p_3))= \hspace{0.5\textwidth}\\
 \frac{1}{2}(A^2_{12,13}+A^2_{13,12}) \pm A_{12,13}A_{13,12} \cos(\delta_{12,13}-\delta_{13,12}) dp, 
\end{multline}
where $p_i$ in parentheses are the momentum of the corresponding particle.
We can then measure $c_i$ using
\begin{equation}
c_i = \frac{1}{2} \left[ \int_i d\Gamma(D_+^0\to\KS(p_1)\pim(p_2)\pip(p_3))
- \int_i d\Gamma(D_-^0\to\KS(p_1)\pim(p_2)\pip(p_3)) \right].
\end{equation}
If we can bin the Dalitz plot so that $c_i$ and $s_i$ are nearly constant
in each bin, $(c_i, s_i)$ can be determined with high precision
\begin{eqnarray}
c_i &=& \sum_j c_{j} = \sum_j A_j A_{\bar j} \cos(\delta_j-\delta_{\bar j})\Delta p_j = \sum_j \sqrt{T_jT_{\bar j}}\cos(\delta_j-\delta_{\bar j}),\\
s_i &=& \sum_j\sqrt{T_jT_{\bar j}}\sin(\delta_j-\delta_{\bar j})
= \sum_j \pm\sqrt{T_jT_{\bar j} - c^2_j}.
\end{eqnarray}
CLEO-c~\cite{Briere:2009aa, Libby:2010nu} measured $s_i$ and $c_i$ for 
$D\to \KS\pip\pim$ and $D\to \KS\Kp\Km$ with 818\invpb of data on $\psi(3770)$
resonance. They also estimated the impact on the measurement of the CKM angle
$\gamma$. They found their $s_i$ and $c_i$ are consistent with that calculated
from the Dalitz plot model used in \babar\ analysis, and the reduction of 
Dalitz plot model dependence is substantial.

\section{Projected precisions in \superb\ era}

In \superb's physics reach studies~\cite{O'Leary:2010af}, the
expected precisions in \Dz-\Dzb mixing parameters in various scenarios on the
\superb\ time scale are estimated.
First the results from \babar's 482\invfb are extrapolated to 
\superb's target of 75\invab near $\Upsilon(4S)$, without any independent 
inputs of strong phase measurements. Then the improvement due to better
precision in strong phase measurements using $D\Db$ threshold data are 
estimated, first from the forthcoming BESIII runs and from 
\superb\ plan (0.5\invab integrated luminosity).

\begin{figure}[htb]
\centering
\includegraphics[width=0.98\textwidth]{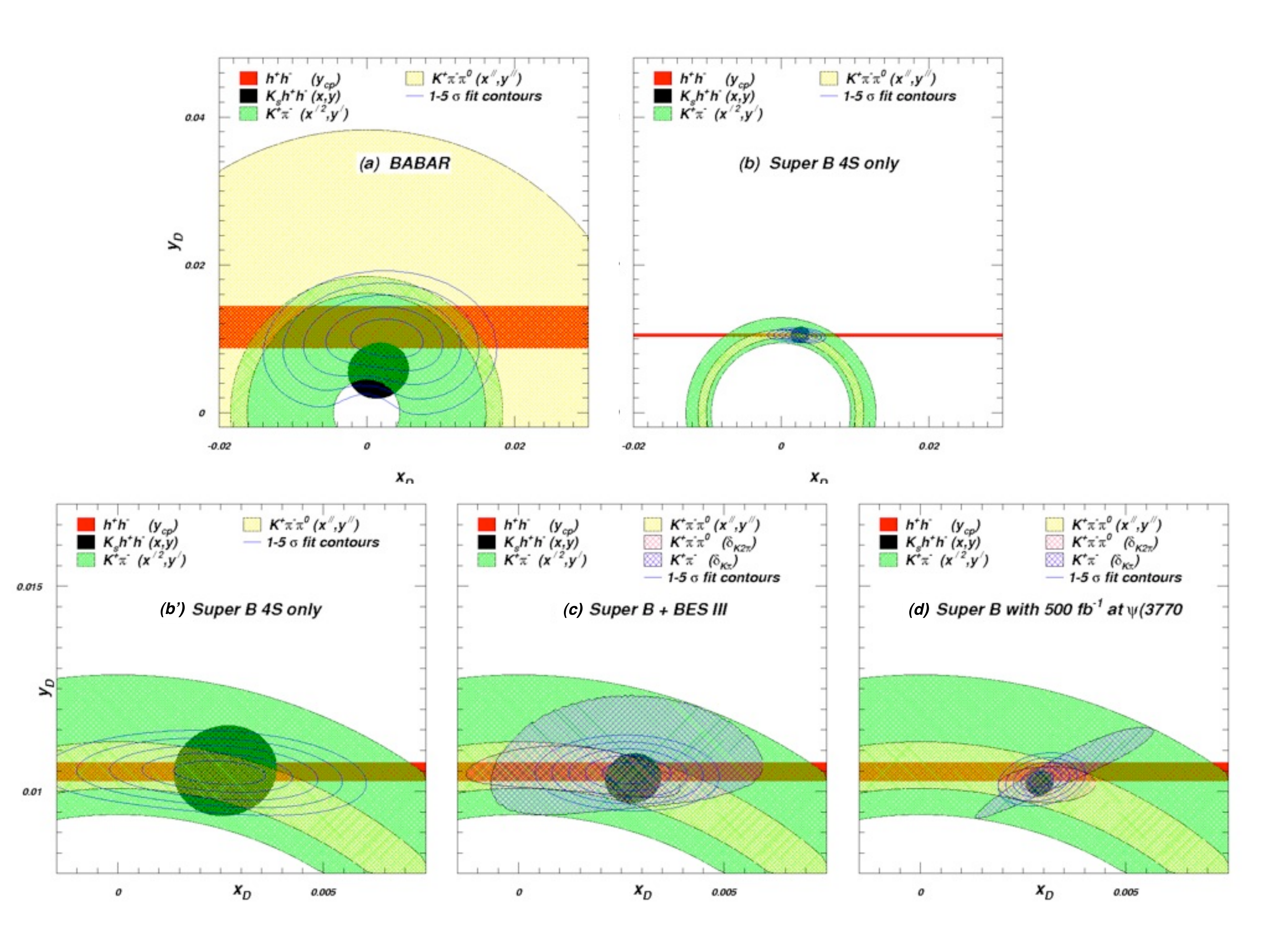}
\caption{
The confidence regions of \Dz-\Dzb mixing parameters $(x,y)$ in various scenarios
described in the text and in Table~\ref{tab:charmproj}.
Shaded areas indicate the coverage of measured observables lying 
   within their 68.3\% confidence region.  Contours enclosing 68.3\% ($1\sigma$), 95.45\% ($2\sigma$), 
   99.73\% ($3\sigma$), 99.994\% ($4\sigma$) and $1-5.7\times 10^{-7}$ 
   two-dimensional confidence regions from the $\chi^2$ fit to these 
   results are drawn as solid lines.
}
\label{fig:charmproj}
\end{figure}

The results, including current average values from \babar, 
are summarize in Table~\ref{tab:charmproj}, and the corresponding
confidence regions are shown in Fig.~\ref{fig:charmproj}.

\begin{table}[t]
\begin{center}
\begin{tabular}{lccccc}
\hline
Fit   & 
$ x \times 10^{3}$                          &
$ y \times 10^{3}$                          &
$\delta_{\Kp\!\pim}^{\circ}$          &
$\delta_{\Kp\!\pim\!\piz}^{\circ}$   \\ [3pt]
\hline
(a) 
  & $  3.01^{+  3.12}_{  -3.39} $
  & $ 10.10^{+  1.69}_{  -1.72} $
  & $ 41.3^{+ 22.0}_{ -24.0}    $
  & $ 43.8  \pm 26.4  $
  \\[3pt]
Stat.
  & $  (2.76) $
  & $  (1.36) $
  & $  (18.8) $
  & $  (22.4) $
  \\[3pt]
(b) 
  & $  xxx ^{+  0.72}_{  -0.75} $
  & $  xxx   \pm 0.19  $
  & $  xxx^{+  3.7}_{  -3.4} $
  & $  xxx^{+  4.6}_{  -4.5} $
\\
Stat.
  & $  (0.18) $
  & $  (0.11) $
  & $  (1.3)  $
  & $  (2.9)  $
  \\[3pt]
(c) 
  & $  xxx  \pm 0.42 $
  & $  xxx   \pm 0.17  $
  & $  xxx  \pm 2.2  $
  & $  xxx^{+  3.3}_{  -3.4} $
\\
Stat.
  & $  (0.18)  $
  & $  (0.11)  $
  & $  (1.3)   $
  & $  (2.7)   $
  \\[3pt]
(d) 
  & $  xxx   \pm 0.20  $
  & $  xxx   \pm 0.12  $
  & $  xxx  \pm 1.0    $
  & $  xxx  \pm 1.1    $
\\
Stat.
  & $  (0.17)  $
  & $  (0.10)  $
  & $  (0.9)   $
  & $  (1.1)   $
 \\
 \hline
\end{tabular}
\caption{\Dz-\Dzb mixing parameters $(x,y)$ and strong phases obtained from
$\chi^2$ fits to observables obtained either from \babar\ or from their
projections to \superb. Fit a) is for $482\invfb$
   from \babar\ alone and this is scaled up in b) to $75\invab$ at
   $\Upsilon(4S)$ for \superb.
   Fit c) includes strong phase information
   projected to come from a BES~III run at $D\bar D$ threshold, and
   d) is what would be possible from a $500\invfb$ $D\bar D$
   threshold run at \superb.  The 
   uncertainties due to statistical limitation alone are shown below each
   fit result.
}
\label{tab:charmproj}
\end{center}
\end{table}

\section{Time-dependent \CP asymmetry}

Using coherent $\psi(3770)\to \Dz\Dzb$ decays, one can perform time-dependent
\CP asymmetry studies analogous to $\Upsilon(4S)\to\Bz\Bzb$ in $B$-factories.
If a neutral $D$ meson decays to a final state at $t_1$ that can identify the 
sign of its $c$-quark, e.g., lepton charge in semileptonic decays, 
the other $D$ meson must be in an
orthogonal state, i.e., the opposite flavor to the first $D$. The time-dependent
decay rate of the second $D$ meson into a \CP eigenstate can be derived from
Eq.~\ref{eq:dGammadt}:
\begin{equation}
A(\Delta t) = \frac{\overline\Gamma(\Delta t)- \Gamma(\Delta t)}{\overline\Gamma(\Delta t)+ \Gamma(\Delta t)} = 2e^{y\Gamma\Delta t}\frac{ (|\lambda_f|^2 -1) \cos (x\Gamma\Delta t) + 2 {\cal I}m\lambda_f \sin(x\Gamma\Delta t)}{(1+|\lambda_f|^2)(1+e^{2y\Gamma\Delta t}) + 2(1-e^{2y\Gamma\Delta t}){\cal R}e\lambda_f},
\end{equation}
where $\Delta t = t_2-t_1$, and $\lambda_f= (q\bar A_f)/(p A_f)$.

\begin{figure}
\centering
\includegraphics[width=0.6\textwidth]{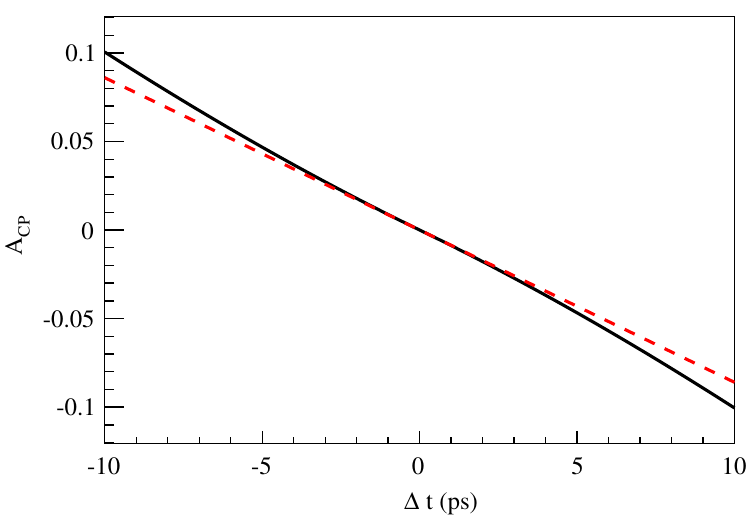}
\caption{Distribution of time-dependent \CP asymmetry for \Dz mesons with
${\cal R}e \lambda_f = {\cal I}m\lambda_f = 1/\sqrt{2}$ and $\CP=-1$. These
distributions assume $q/p=1$ and the solid (dashed) line corresponds to $y=0.01 (0.00).$}
\label{fig:tdcp-charm}
\end{figure}

Measuring time-dependent \CP asymmetry in \Dz-\Dzb system is much more difficult
than in \Bz-\Bzb system. The reason is that charm mixing rate is very small; 
both $x$ and $y$ are ${\cal O}(1\%)$ for \Dz-\Dzb, whereas $x\sim {\cal O}(1)$ for
\Bz-\Bzb. This effect is illustrated in Fig.~\ref{fig:tdcp-charm}~\cite{Bevan:2011up}, in which 
one can see that even with a large \CP-violating phase 
(${\rm arg}(\lambda_f)= \pi/4$) the \CP asymmetry is only a few percent
within $|\Delta t|<10$~ps (more than 20 times the \Dz lifetime). 
In contrast, within the same $\Delta t$ range, 
the \CP asymmetry for \Bz meson exhibits 1.5 full sinusoidal oscillations 
already.

\begin{figure}
\centering
\includegraphics[width=0.99\textwidth]{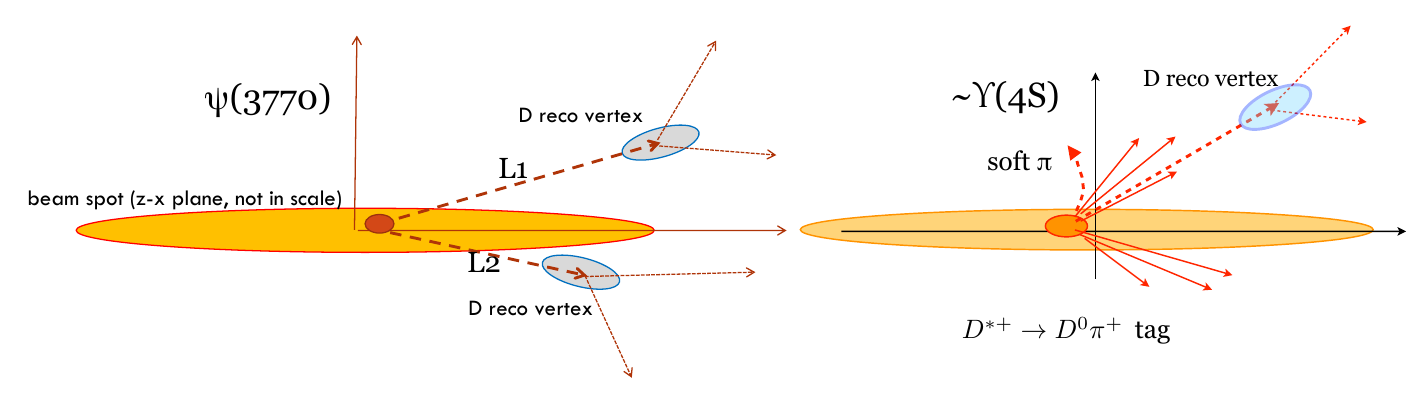}
\caption{Illustrations of charm meson reconstructions with beam spot at 
\superb\
for (left) $\psi(3770)\to D\Db$ events and for (right) continuum events 
near $\Upsilon(4S)$.}
\label{fig:beamspot}
\end{figure}

At \superb, the design beam spot is much smaller ($\sigma_x\sim 8~\mu{\rm m}$, 
$\sigma_y\sim 40~{\rm nm}$, $\sigma_z\sim 200~\mu{\rm m}$) than that in \babar.
This makes fitting for the primary vertex possible (and meaningful) even if
no charged tracks originating from the primary vertex. As illustrated in 
Fig.~\ref{fig:beamspot}, the charm mesons from $\psi(3770)$ decay fly away
from the primary vertex for ${\cal O}(100~\mu{\rm m})$, depending on the
center-of-mass frame boost. One can perform a beam-spot constraint fit on the
$\psi(3770)\to D \Db$ system to simultaneously fit for both flight lengths 
($L_1$ and $L_2$) and convert them to decay times. At near $\Upsilon(4S)$,
one studies charm physics using continuum data; the soft pion from $D^*$ 
decays are used to identify the initial flavor of the charm meson.

\superb\ has conducted studies to evaluate the sensitivities to mixing 
parameters $(x,y)$ and \CP-violating parameters $q/p$ using 0.5\invab
of $D\Db$ threshold data alone and compared that with using 75\invab of
data near $\Upsilon(4S)$. The preliminary results that used several two-body
charm decays with various combination of \CP/flavor-tags shows that the
uncertainties with $\psi(3770)$ data are about six times larger than those 
with $\Upsilon(4S)$ data. On this topic along, 0.5\invab of $D\Db$ threshold 
data is clearly not as competitive as $\Upsilon(4S)$ data. However, one should 
be reminded that the former only requires a few months of data taking, while
the latter will take five years according to the plan.

\section{Summary}

The precision of charm mixing measurements will be limited by the uncertainties
of strong phases and Dalitz plot model by the time \superb\ collected its
targeted data near $\Upsilon(4S)$.
One can mitigate this situation by
utilizing the quantum correlation of charm decays in $\psi(3770)\to D\Db$
with BESIII data. 
With a months-long run at $D\Db$ threshold at \superb, it is possible to
improve the precision by another factor of two.
With a boost of the center-of-mass frame, time-dependent \CP\ asymmetry 
measurements can also be performed in $\psi(3770)\to D\Db$ data, but the
precision is not as competitive as the much larger $\Upsilon(4S)$ data.

Finally, not discussed in this paper but worth noting here, charm threshold 
data have advantages to $\Upsilon(4S)$ data in several areas due to the low
background and the fact that the whole event can be fully reconstructed
(double tag), in addition to the quantum correlation. These areas include
rare decays ($\Dz\to\gamma\gamma,\ \mu\mu(X)$, etc.), leptonic/semileptonic
charm decays, form factor measurements, \CPT\ violation, \CP\ violation in 
$D\to V\gamma$ that probes chromomagnetic dipole 
operator~\cite{Isidori:2012yx}, and others. 
It certainly adds to the breadth of \superb\ physics programs.


\begin{thebibliography}{99}

\bibitem{O'Leary:2010af} 
  B.~O'Leary {\it et al.}  [\superb\ Collaboration],
  [arXiv:1008.1541 [hep-ex]].

\bibitem{Amhis:2012bh} 
  Y.~Amhis {\it et al.}  [Heavy Flavor Averaging Group Collaboration],
  [arXiv:1207.1158 [hep-ex]].

\bibitem{Burdman:2003rs} 
  See, for example, G.~Burdman and I.~Shipsey,
  Ann.\ Rev.\ Nucl.\ Part.\ Sci.\  {\bf 53}, 431 (2003)
  [hep-ph/0310076] and references therein.

\bibitem{Asner:2008nq} 
  D.~M.~Asner, T.~Barnes, J.~M.~Bian, I.~I.~Bigi, N.~Brambilla, I.~R.~Boyko, V.~Bytev and K.~T.~Chao {\it et al.},
  Int.\ J.\ Mod.\ Phys.\ A {\bf 24}, S1 (2009)
  [arXiv:0809.1869 [hep-ex]].

\bibitem{Asner:2008ft} 
  D.~M.~Asner {\it et al.}  [CLEO Collaboration],
  Phys.\ Rev.\ D {\bf 78}, 012001 (2008)
  [arXiv:0802.2268 [hep-ex]].

\bibitem{Asner:2005sz} 
  D.~M.~Asner {\it et al.}  [CLEO Collaboration],
  Phys.\ Rev.\ D {\bf 72}, 012001 (2005)
  [hep-ex/0503045].


\bibitem{Abe:2007rd} 
  K.~Abe {\it et al.}  [BELLE Collaboration],
  Phys.\ Rev.\ Lett.\  {\bf 99}, 131803 (2007)
  [arXiv:0704.1000 [hep-ex]].

\bibitem{delAmoSanchez:2010xz} 
  P.~del Amo Sanchez {\it et al.}  [\babar\ Collaboration],
  Phys.\ Rev.\ Lett.\  {\bf 105}, 081803 (2010)
  [arXiv:1004.5053 [hep-ex]].

\bibitem{Giri:2003ty} 
  A.~Giri, Y.~Grossman, A.~Soffer and J.~Zupan,
  Phys.\ Rev.\ D {\bf 68}, 054018 (2003)
  [hep-ph/0303187].

\bibitem{Bondar:2010qs} 
  A.~Bondar, A.~Poluektov and V.~Vorobiev,
  Phys.\ Rev.\ D {\bf 82}, 034033 (2010)
  [arXiv:1004.2350 [hep-ph]].

\bibitem{Briere:2009aa} 
  R.~A.~Briere {\it et al.}  [CLEO Collaboration],
  Phys.\ Rev.\ D {\bf 80}, 032002 (2009)
  [arXiv:0903.1681 [hep-ex]].

\bibitem{Libby:2010nu} 
  J.~Libby {\it et al.}  [CLEO Collaboration],
  Phys.\ Rev.\ D {\bf 82}, 112006 (2010)
  [arXiv:1010.2817 [hep-ex]].

\bibitem{Bevan:2011up} 
  A.~J.~Bevan, G.~Inguglia and B.~Meadows,
  Phys.\ Rev.\ D {\bf 84}, 114009 (2011)
  [arXiv:1106.5075 [hep-ph]].

\bibitem{Isidori:2012yx} 
  G.~Isidori and J.~F.~Kamenik,
  arXiv:1205.3164 [hep-ph].

\end{thebibliography}
\end{document}